\documentstyle[aps,pre]{revtex}

\begin{document}

\title{Detailed balance and $H$-theorems for dissipative particle dynamics}
\author{C. A. Marsh \footnote{email:c.marsh1@physics.ox.ac.uk} \\
  Theoretical Physics, Keble Road, Oxford.  OX1 3NP.  UK\\
  P. V. Coveney \footnote{email:coveney@cambridge.scr.slb.com} \\
  Theoretical Physics, Keble Road, Oxford.  OX1 3NP.  UK and\\
  Sclumberger Cambridge Research, High Cross, Madingley Road, 
  Cambridge, CB3 OE2. UK}

\maketitle
\begin{center}{\bf Abstract}\end{center}
\begin{abstract}
An extension of the $H$-theorem for dissipative particle dynamics
(DPD) to the case of a multi-component fluid is made.  Detailed balance and an
additional $H$-theorem are proved for an energy-conserving version of the 
DPD algorithm.  The implications of these results for the 
statistical mechanics of the method are discussed.
\end{abstract}



\newcommand{\wfd}{w_{\scriptscriptstyle D}}
\newcommand{\wfr}{w_{\scriptscriptstyle R}}
\newcommand{\wfk}{A_{\scriptscriptstyle D}}
\newcommand{\wfe}{A_{\scriptscriptstyle R}}
\newcommand{\ndsum}{\sum_{j \neq i}}

\section{Introduction}

Interest in the rheological and dynamical properties of complex 
fluids~\cite{McL}
over the past decade has seen the introduction of several new
techniques for their simulation on {\em mesoscopic} length scales.
These methods include lattice gas automata (LGA), 
lattice Boltzmann equation (LBE) and dissipative particle dynamics
(DPD)~\cite{BAC}.  The aim of this letter is to explore 
some of the statistical mechanical properties of the rapidly evolving 
DPD model, and to extend these results in order to keep
pace with new developments being made in the algorithms.

The DPD method was originally introduced by Hoogerbrugge and Koelman
\cite{HK} as a discrete time algorithm; this was subsequently modified and
reinterpreted as a discrete time approximation to an underlying system
obeying Langevin dynamics (with momentum conservation) by Espa\~nol
and Warren \cite{EW}, in order
to guarantee the existence of a Gibbsian (specifically a canonical) 
equilibrium state.  Applications of the model include
colloidal suspensions \cite{colloids}, polymer suspensions
\cite{polymers} and binary mixtures \cite{CN,NC}.  A dynamical theory
has been presented \cite{MBE1,MBE2} for the continuous time limit
and the equilibrium for finite time-step has been investigated \cite{MY}.

We briefly describe the implementation of the DPD method.  The system
consists of a set of $N$ discrete particles which move in continuous
space and in discrete time-steps, the interval between which may be
reduced to being infinitesimal.  At each time-step $\delta t$, 
the particles' momenta are updated by a momentum-conserving interaction with 
each particle inside a neighbourhood of radius $R_0$.  
This interaction 
includes three distinct forces, which
can be described as {\em conservative} ${\bf F}^C$,
{\em dissipative} ${\bf F}^D$ and {\em random} ${\bf F}^R$. 
Between each tick of the clock, 
the particles all propagate freely according to
their velocities.  In the limit of continuous time, 
the DPD equations of motion are most effectively
described in terms of the following stochastic differential equations:

\begin{eqnarray}
\dot{\bf v}_i &=& \ndsum \left\{
{\bf F}_{ij}^C + {\bf F}_{ij}^D + {\bf F}_{ij}^R \right\} \\
\dot{\bf r}_i &=& {\bf v}_i.
\end{eqnarray}

\noindent for each particle labelled by the subscript $i$. 
The forces take the following forms:

\begin{eqnarray}
{\bf F}_{ij}^C &=& -\frac{1}{m} \frac{\partial \phi}{\partial r_{ij}} \\
{\bf F}_{ij}^D &=& -\gamma \wfd(r_{ij}) [ {\bf e}_{ij} \cdot
        {\bf v}_{ij} ] {\bf e}_{ij} \\
{\bf F}_{ij}^R &=& \sigma \wfr(r_{ij}) {\bf e}_{ij} \zeta_{ij}
\end{eqnarray}

\noindent 
where $\phi$ is a potential energy, ${\bf r}_{ij}={\bf r}_i-{\bf r}_j$
is the relative separation vector and ${\bf e}_{ij}$ is the unit
vector in the direction of ${\bf r}_{ij}$; for simplicity, all
particles are assumed here to have the same mass $m$.  The functions
$\wfd(r_{ij})$ and $\wfr(r_{ij})$ are weighting functions which limit the
action of the dissipative and random forces to a finite range $R_0$.
The random elements $\zeta_{ij}$ are Gaussian white noise with zero
mean: $\overline{\zeta_{ij}}=0$.  They are uncorrelated for
different pairs of particles and for different times:
$\overline{\zeta_{ij}(t)\zeta_{kl}(t')}=
(\delta_{ik} \delta_{jl} + \delta_{il} \delta_{jk}) \delta(t-t')$.
It should be noted that these forces conserve both momentum and
angular momentum but not energy.

After a brief description of the statistical mechanical concepts
involved, we present an $H$-theorem for the multi-component DPD fluid.
We then briefly describe the energy conserving DPD model and examine
detailed balance and an $H$-theorem in this context.

\section{On detailed balance and $H$-theorems} It is important at the
outset to explain the significance and importance of the statistical
mechanical properties of detailed balance and the so-called
$H$-theorem. Detailed balance is known to be a sufficient, but not
necessary, condition which ensures that a Gibbsian equilibrium state
exists in the ensemble representation of a dynamical system. It is
possible for systems not satisfying detailed balance to exhibit
equilibrium states; however, characterising these is then a much
harder task. The virtue of the modifications made by
Espa\~nol and Warren to the original DPD algorithm is that, in the
limit of continuous time, the $N$-body DPD system then satisfies the
detailed balance condition, thereby guaranteeing the existence of a well
defined equilibrium state.

In the literature, at least two separate kinds of ``$H$-theorem'' can
be distinguished. First, any Markov chain or process which has an
equilibrium distribution will have an $H$-theorem associated with it,
in the sense of possessing a Lyapounov function that changes
monotonically with time. Indeed, a
whole class of Lyapounov functions achieve this; the class is defined as
the expectation of any convex function of the relative-to-equilibrium
probability density. The proof of such $H$-theorems follows directly
from the linear equation (also referred to as the ``master equation'')
for the $N$-body distribution function. Detailed balance plays a role
here, in that it specifies what the equilibrium distribution is; this
information is needed to write down the Lyapounov function.

However, the arguably more famous Boltzmann $H$-theorem 
is quite a different notion, and is of much more restricted validity.
Boltzmann's $H$ function is defined in terms of the {\em one}-body
distribution function, and its time-monotonicity can only be derived if
we know the kinetic equation obeyed by this reduced distribution.
Moreover, this equation is non-linear so that the choice of $H$ is now
much more restricted; for the Boltzmann equation itself, only the 
expectation of the logarithm of this reduced probability distribution 
suffices. In proving Boltzmann's $H$-theorem, use is made of the 
property of detailed balance.

Note in passing that time-symmetry is a stronger property than detailed
balance; the former implies the latter, but is not implied by it.
Thus, detailed balance is obeyed by both Newton's equations of motion
and by dissipative particle dynamics, although the former is
time-symmetric while the latter is not.

The existence of an $H$-theorem for a given system can be used to
check on the numerical stability of any algorithm implemented to
simulate it; numerical instabilities which lead to non-monotonicity of
the $H$-functional concerned 
can then be precluded. The issue of the existence of 
detailed balance and related $H$-theorems
is thus clearly of importance for the various mesoscale modelling and
simulation techniques.

By contrast with (continuous time) DPD, 
virtually all interacting lattice-gas and lattice-Boltzmann models
have no known detailed balance condition; therefore, their equilibrium
states are generally unknown, while the lack of any associated
$H$-theorems makes the real-valued lattice-Boltzmann methods, 
in particular, 
susceptible to poorly understood numerical instabilities. Indeed,
because detailed balance is not satisfied in such models, it makes
their theoretical analysis by standard methods of non-equilibrium
statistical mechanics well nigh impossible.

\section{$H$-theorem for multicomponent, isothermal DPD} Detailed
balance and an $H$-theorem (of the first kind mentioned in the
preceding section, i.e. for the full $N$-body distribution) 
for the single component DPD fluid have already been
demonstrated \cite{MBE1}.  The proof of detailed balance
for general DPD models of interacting multi-component 
fluids has also been derived \cite{DB}.
Here, we aim to extend this form of $H$-theorem to the case of a
multi-component fluid, which includes the case of binary immiscible
fluids~\cite{CN,NC}.  

It has been demonstrated \cite{DB} that 
the evolution equation for the $N$-particle distribution
function is the Fokker-Planck equation for the multi-component fluid:
$\partial_t P = {\cal L}^{\rm \scriptscriptstyle MC} P$,
where the multi-component Fokker-Planck operator
${\cal L}^{\rm \scriptscriptstyle MC}$ is defined as:

\begin{eqnarray} \label{MC_FP}
{\cal L}^{\rm \scriptscriptstyle MC} &=&
        -\left[ \sum_{\alpha} \sum_{i_\alpha} {\bf v}_{i_\alpha} \cdot
        \frac{\partial}{\partial {\bf r}_{i_\alpha}} +
        \sum_{\alpha \beta} \sum_{i_\alpha j_\beta}
        \frac{{\bf F}_{i_\alpha j_\beta}^C}{m} \cdot \frac{\partial}
        {\partial {\bf v}_{i_\alpha}} \right] \\
&+& \sum_{\alpha \beta} \sum_{i_\alpha j_\beta} \frac{\gamma}{m} w_D
        (r_{i_\alpha j_\beta}) \left[ {\bf e}_{i_\alpha j_\beta} \cdot
        \frac{\partial}{\partial {\bf v}_{i_\alpha}} \right]
        {\bf e}_{i_\alpha j_\beta} \cdot 
        \left[ {\bf v}_{i_\alpha j_\beta}
        +\frac{\theta}{m} \left( \frac{\partial}
        {\partial {\bf v}_{i_\alpha}} - \frac{\partial}
        {\partial {\bf v}_{j_\beta}} \right) \right] \nonumber
\end{eqnarray}
 
\noindent where $\alpha$ and $\beta$ are sums over different types of
particles and $i_\alpha$ and $j_\beta$ sum over all particles of each
type.  The parameter $\theta$ is defined as $\theta=m\sigma^2/2\gamma$.
The relevant $H$-functional for the multi-component case is a simple
extension of that in the single component case.  As expected for an
isothermal system, it is just the expectation of the associated free
energy $<U-\theta S>$, where $U$ is the internal energy, $\theta$ is the
equilibrium temperature, $S$ is the global entropy, and the
expectation is taken using the full $N$-particle distribution function, $P$:

\begin{equation}
{\cal F}[P({\bf \Gamma},t)] = \int d{\bf \Gamma} P
\left\{ \sum_\alpha \sum_{i_\alpha} 
\left[ \frac{mv_{i_\alpha}^2}{2} +
\sum_\beta \sum_{j_\beta} 
V(r_{i_\alpha j_\beta}) \right] + \theta \ln
P \right\}
\end{equation}

Using the time evolution operator for the multi-component system
(\ref{MC_FP}), it is possible to show that:

\begin{equation}
\frac{d{\cal F}}{dt} = -\sum_{\alpha \beta}
        \sum_{i_\alpha j_\beta} \int d{\bf \Gamma}
        \frac{\gamma w_D(r_{i_\alpha j_\beta})}{P} \left[
        {\bf e}_{i_\alpha j_\beta} \cdot \left[
        {\bf v}_{i_\alpha j_\beta} + \frac{\theta}{m} 
        \left( \frac{\partial}{\partial {\bf v}_{i_\alpha}} -
        \frac{\partial}{\partial {\bf v}_{j_\beta}} \right) \right] P
        \right]^2
\end{equation}

It is then apparent that the time derivative of the functional 
${\cal F}$ is the sum of negative definite terms, and therefore 
that the functional itself is monotonically decreasing in time.
The appropriate equilibrium distribution for the multi-component
system occurs when this functional stops decreasing.  It is easy to
show that this occurs when it reaches the Gibbsian distribution for 
the associated conservative system, i.e. as if the dissipative and
random forces were not present:

\begin{equation}
P^{\rm \scriptscriptstyle}_{eqm} = \frac{1}{Z_{\scriptscriptstyle MC}} 
        \exp \left\{ -\frac{1}{\theta}
        \sum_\alpha \sum_{i_\alpha} 
        \left[ \frac{mv_{i_\alpha}^2}{2} +
        \sum_\beta \sum_{j_\beta} 
        V(r_{i_\alpha j_\beta}) \right] \right\},
\end{equation}

\noindent $Z_{\scriptscriptstyle MC}$ being the multi-component 
canonical partition function, defined in the normal way.

\section{Energy Conserving DPD} An energy conserving version of 
DPD has recently been presented by 
Espa\~nol~\cite{energy}.  This involves the introduction of an internal energy
variable $\epsilon_i$ for each DPD particle (now interpreted as a cluster
of atoms or molecules, into which the dissipated energy is assumed to
flow).  There is an entropy
$s(\epsilon_i)$ which needs to be specified in order to 
describe a given system, and the
temperature is defined in the usual thermodynamic way as 
$\theta_i = (\partial s_i/\partial \epsilon_i)^{-1}$.
It is then possible to formulate a new DPD algorithm which conserves 
the total energy of the system, as well as momentum and angular
momentum~\cite{energy}.  
An appropriate set of stochastic differential equations is:

\begin{eqnarray}
\dot{\bf r}_i &=& {\bf v}_i \\
\dot{\bf v}_i &=& \ndsum \left[
        \frac{1}{m}F_{ij}^C - \gamma_{ij} \wfd(r_{ij}) ({\bf e}_{ij} \cdot
        {\bf v}_{ij}) {\bf e}_{ij} + \sigma_{ij} \wfr(r_{ij}) {\bf e}_{ij}
        \zeta_{ij} \right] \\
\dot{\bf \epsilon}_i &=& \frac{m}{2} \ndsum \left[
        \gamma_{ij} \wfd(r_{ij}) ({\bf v}_{ij} \cdot {\bf e}_{ij})^2 -
        \sigma_{ij}^2 \wfr^2(r_{ij}) - \sigma_{ij} \wfr(r_{ij})
        ({\bf e}_{ij} \cdot {\bf v}_{ij}) \zeta_{ij} \right. \nonumber \\
&&+ \left. \kappa_{ij} \left\{ \frac{1}{\theta_i} - \frac{1}{\theta_j} \right\}
        \wfk(r_{ij}) + \alpha_{ij} \wfe(r_{ij}) \zeta_{ij}^{\epsilon}
        \right]
\end{eqnarray}  

\noindent Here the functions
$\wfk(r_{ij})$ and $\wfe(r_{ij})$ are additional weighting functions
for what can be interpreted as the {\em conduction} and {\em random
heat flux} terms respectively, while $\kappa_{ij}$ and $\alpha_{ij}$ are
their strengths;
$\zeta_{ij}^{\epsilon}$ are random elements which are
uncorrelated to the elements $\zeta_{ij}$ and have zero mean 
$\overline{\zeta_{ij}^\epsilon}=0$.  They are uncorrelated for different times
and different pairs of particles and are anti-symmetric:
$\overline{\zeta_{ij}^\epsilon(t) \zeta_{kl}^\epsilon(t')}=(\delta_{ik}
\delta_{jl}-\delta_{il} \delta_{jk}) \delta(t-t')$.  
We also note that the strengths of the random and 
dissipative forces ($\sigma$ and $\gamma$)
can now, in general, vary for different particle pairs.

Following the original derivation~\cite{energy}, we make the 
additional assumptions:

\begin{equation}
\wfk^2(r) = \wfe(r), \qquad \alpha_{ij}^2 = 2\kappa_{ij}, \qquad
\wfr^2(r) = \wfd(r).
\end{equation}

\noindent which mean that the Fokker-Planck equation for the evolution of the
$N$-particle distribution function $P$, can be written as:

\begin{equation} \label{FP}
\partial_t P = \left[ {\cal L}_C + {\cal L}_{VH} + {\cal L}_{HC} \right] P
\end{equation}

\noindent
where the operators on the right hand side are defined as follows:

\begin{eqnarray}
{\cal L}_C &=& - \sum_i {\bf v}_i \cdot 
        \frac{\partial}{\partial {\bf r}_i}
        - \ndsum \frac{{\bf F}_{ij}^C}{m} \cdot
        \frac{\partial}{\partial {\bf v}_i} \\
{\cal L}_{VH} &=& \frac{1}{2} \ndsum \wfd(r_{ij}) L_{ij} \left[
        \gamma_{ij} ({\bf v}_{ij} \cdot {\bf e}_{ij}) + L_{ij}
        \frac{\sigma_{ij}^2}{2} \right] \\
{\cal L}_{HC} &=& \ndsum \wfk(r_{ij}) \frac{\partial}
        {\partial \epsilon_i} \left[ \frac{1}{\theta_j} - \frac{1}{\theta_i}
        + \frac{\partial}{\partial \epsilon_i} 
        - \frac{\partial}{\partial \epsilon_j} \right]
        \kappa_{ij} \\
L_{ij} &=& {\bf e}_{ij} \cdot \left[ \frac{\partial}{\partial {\bf v}_i}
        - \frac{\partial}{\partial {\bf v}_j} - \frac{m}{2} {\bf v}_{ij}
        \left[ \frac{\partial}{\partial \epsilon_i} +
        \frac{\partial}{\partial \epsilon_j} \right] \right]
\end{eqnarray}

where the subscripts $C$, $VH$ and $HC$ refer to the {\em Conservative},
{\em Viscous Heating} and {\em Heat Conduction} terms respectively.

\section{Detailed balance for energy-conserving DPD}
If the evolution operator for a system is ${\cal L}$ and we designate
its adjoint by the operator ${\cal L}^\dagger$, then the detailed
balance constraint \cite{RISKEN} can be written in the following way:

\begin{equation} \label{db}
{\cal L} P_{eqm} \varphi = P_{eqm} {\cal L}^{\dagger \epsilon} \varphi
\end{equation}

\noindent where the superscript $\epsilon$ indicates that all 
variables that are {\em odd} under time reversal are to have their
signs reversed.  In the case of DPD, this means that the velocities
attract an additional minus sign.  The function $\varphi$ can be any
function of the phase space variables.

The appropriate operators for the energy-conserving version of DPD are:

\begin{eqnarray}
{\cal L}_C^{\dagger \epsilon} 
        &=& - \sum_i {\bf v} \cdot \frac{\partial}{\partial {\bf r}_i}
        - \ndsum \frac{{\bf F}_{ij}^C}{m} \cdot
        \frac{\partial}{\partial {\bf v}_i} = {\cal L}_C \\
{\cal L}_{VH}^{\dagger \epsilon} 
        &=& \frac{1}{2} \ndsum \wfd(r_{ij}) \left[
        - \gamma_{ij} ({\bf v}_{ij} \cdot {\bf e}_{ij}) + L_{ij}
        \frac{\sigma_{ij}^2}{2} \right] L_{ij} \\
{\cal L}_{HC}^{\dagger \epsilon} 
        &=& - \ndsum \wfk(r_{ij}) \kappa_{ij}
        \left[ \frac{1}{\theta_j} - \frac{1}{\theta_i}
        + \frac{\partial}{\partial \epsilon_i} 
        - \frac{\partial}{\partial \epsilon_j} \right]
        \frac{\partial} {\partial \epsilon_i} 
\end{eqnarray}

We can then show that:

\begin{eqnarray} \label{LC_DB}
{\cal L}_C P_{eqm} \varphi &=& P_{eqm} {\cal L}_C \varphi 
        + \varphi {\cal L}_C P_{eqm} \nonumber \\
&=& P_{eqm} {\cal L}_C^{\dagger \epsilon} \varphi,
\end{eqnarray}

\begin{eqnarray} \label{LVH_DB}
{\cal L}_{VH} P_{eqm} \varphi &=& 
        \frac{1}{2} \ndsum \wfd(r_{ij}) L_{ij} \left\{
        \gamma_{ij} ({\bf e}_{ij} \cdot {\bf v}_{ij}) + L_{ij}
        \frac{\sigma^2}{2} \right\} P_{eqm} \varphi \nonumber \\
&=&     \frac{1}{2} \ndsum \wfd(r_{ij}) L_{ij} P_{eqm} L_{ij}
        \frac{\sigma^2}{2} \varphi \nonumber \\
&=&     \frac{1}{2} \ndsum \wfd(r_{ij}) L_{ij} P_{eqm} \left[
        -\gamma_{ij} ({\bf e}_{ij} \cdot {\bf v}_{ij}) + L_{ij}
        \frac{\sigma^2}{2} \right] L_{ij} \varphi \nonumber \\
&=&     P_{eqm} {\cal L}_{VH}^{\dagger \epsilon},
\end{eqnarray}

and

\begin{eqnarray} \label{LHC_DB}
{\cal L}_{HC} P_{eqm} \varphi &=& 
        \ndsum \wfk(r_{ij}) \frac{\partial}{\partial \epsilon_i} \left[
        \frac{1}{\theta_i} - \frac{1}{\theta_j} + \frac{\partial}
        {\partial \epsilon_i} - \frac{\partial}{\partial \epsilon_j}
        \right] \kappa_{ij} P_{eqm} \varphi \nonumber \\
&=&     \frac{1}{2} \ndsum \wfk(r_{ij}) \kappa_{ij}
        \left[ \frac{\partial}{\partial \epsilon_i} -
        \frac{\partial}{\partial \epsilon_j} \right] \left[
        \frac{1}{\theta_i} - \frac{1}{\theta_j} + \frac{\partial}
        {\partial \epsilon_i} - \frac{\partial}{\partial \epsilon_j}
        \right] P_{eqm} \varphi \nonumber \\
&=&     \frac{1}{2} \ndsum \wfk(r_{ij}) \kappa_{ij}
        \left[ \frac{\partial}{\partial \epsilon_i} -
        \frac{\partial}{\partial \epsilon_j} \right] P_{eqm}
        \left[ \frac{\partial}{\partial \epsilon_i} -
        \frac{\partial}{\partial \epsilon_j} \right] \varphi
        \nonumber \\
&=&     \frac{1}{2} \ndsum \wfk(r_{ij}) \kappa_{ij} P_{eqm}
        \left[ \frac{1}{\theta_i} - \frac{1}{\theta_j} + \frac{\partial}
        {\partial \epsilon_i} - \frac{\partial}{\partial \epsilon_j}
        \right] \left[ \frac{\partial}{\partial \epsilon_i} -
        \frac{\partial}{\partial \epsilon_j} \right] \varphi
        \nonumber \\
&=&     \ndsum \wfk(r_{ij}) \kappa_{ij} P_{eqm}
        \left[ \frac{1}{\theta_i} - \frac{1}{\theta_j} + \frac{\partial}
        {\partial \epsilon_i} - \frac{\partial}{\partial \epsilon_j}
        \right] \frac{\partial}{\partial \epsilon_i} \varphi \nonumber
        \\
&=&     P_{eqm} {\cal L}_{HC}^{\dagger \epsilon} \varphi,
\end{eqnarray}

\noindent
where we choose the strengths of the dissipative and random
forces to satisfy the following relations:

\begin{equation}
\sigma_{ij} = \sigma, \qquad \gamma_{ij} = \frac{m\sigma^2}{4}
\left[ \frac{1}{\theta_i} + \frac{1}{\theta_j} \right].
\end{equation}

It is therefore apparent that:

\begin{equation}
[ {\cal L}_C + {\cal L}_{VH} + {\cal L}_{HC} ] P_{eqm} \varphi =
P_{eqm} [ {\cal L}_C^{\dagger \epsilon} + {\cal L}_{VH}^{\dagger
\epsilon} + {\cal L}_{HC}^{\dagger \epsilon} ] \varphi
\end{equation}

\noindent
and therefore that the energy-conserving DPD algorithm satisfies detailed
balance.   The equilibrium distribution associated with this detailed
balance condition satisfies the following relations:

\begin{eqnarray}
\left[ L_{ij} + ({\bf e}_{ij} \cdot {\bf v}_{ij}) \frac{m}{2} \left(
        \frac{1}{\theta_i} + \frac{1}{\theta_j} \right) \right] P_{eqm} &=& 0
        \label{DB1}\\
\left[ \frac{1}{\theta_j} - \frac{1}{\theta_i} + 
        \frac{\partial}{\partial \epsilon_i} - 
        \frac{\partial}{\partial \epsilon_j} \right] P_{eqm} &=& 0. \label{DB2}
\end{eqnarray}

\noindent Therefore the equilibrium distribution consistent with
$[{\cal L}_C + {\cal L}_{VH} + {\cal L}_{HC}]P_{eqm}=0$ has the
following form:

\begin{equation}
P_{eqm} = \frac{1}{Z_{\scriptscriptstyle EC}} 
\exp \left\{ \sum_i s_i(\epsilon_i) \right\}
\end{equation}

\noindent 
where $Z_{\scriptscriptstyle EC}$ is the normalization constant.

\section{$H$-theorem for energy-conserving DPD}
The $H$-theorem for energy-conserving DPD can now be formulated.  
We first define the following $H$-functional:

\begin{equation} \label{functional}
S[P({\bf \Gamma})](t) = \int d{\bf \Gamma} P({\bf \Gamma},t) \left[
        \left\{ \sum_i s_i \right\} - \ln P({\bf \Gamma},t) \right]
\end{equation}

\noindent Then, with the aid of the appropriate 
Fokker-Planck equation (\ref{FP}), it is
possible to show that the time-evolution of this functional is:

\begin{eqnarray}
\frac{d S[P({\bf \Gamma})](t)}{dt} &=& \frac{\sigma^2}{4} \int 
        d{\bf \Gamma} \ndsum \frac{\wfd(r_{ij})}{P} \left[
        \left\{ L_{ij} + ({\bf e}_{ij} \cdot {\bf v}_{ij}) \frac{m}{2}
        \left( \frac{1}{\theta_i} + \frac{1}{\theta_j} \right) 
        \right\} P \right]^2
        \nonumber \\
&&      + \frac{1}{2} \int d{\bf \Gamma} \ndsum 
        \frac{\wfk(r_{ij})}{P}
        \left[ \left\{ \frac{1}{\theta_j} - \frac{1}{\theta_j} + 
        \frac{\partial}{\partial \epsilon_i} -
        \frac{\partial}{\partial \epsilon_j} \right\} P
        \right]^2 \kappa_{ij}
\end{eqnarray}

\noindent We note that the time derivative of the functional $S$
consists of sums of two types of terms, each of which is 
positive definite.  Therefore $S[P({\bf \Gamma})]$ is 
monotonically increasing in time and 
the equilibrium is reached when this time evolution stops.  It is easy
to show that this can only be achieved when the equilibrium
distribution satisfies the following relations:

\begin{eqnarray}
\left[ L_{ij} + ({\bf e}_{ij} \cdot {\bf v}_{ij}) \frac{m}{2} \left(
        \frac{1}{\theta_i} + \frac{1}{\theta_j} \right) \right] 
        P_{eqm} &=& 0 \\
\left[ \frac{1}{\theta_j} - \frac{1}{\theta_i} + 
        \frac{\partial}{\partial \epsilon_i} - 
        \frac{\partial}{\partial \epsilon_j} \right] P_{eqm} &=& 0
\end{eqnarray}

\noindent
and therefore that the equilibrium distribution consistent with
$[{\cal L}_C + {\cal L}_{VH} + {\cal L}_{HC}]P_{eqm}=0$ will be:

\begin{equation} \label{EQM_EC}
P_{eqm} = \frac{1}{Z_{\scriptscriptstyle EC}} 
        \exp \left\{ \sum_i s_i(\epsilon_i) \right\}
\end{equation}

\noindent 
This is naturally the same equilibrium distribution
already recognised as being a stationary point of the Fokker-Planck
evolution~eqns~(\ref{DB1}) and (\ref{DB2}). The $N$-body $H$-theorem
provides additional information because it guarantees that the system
approaches this equilibrium state monotonically.  

The $H$-functional~(\ref{functional}) may be interpreted as 
the total entropy of the system.  We see that the first term in the
functional represents the
microscopic entropy of each DPD particle while the second term represents
the normal macroscopic entropy $-P\ln P$.  That the relevant functional is in
this case the total system entropy, rather than a free energy, 
could be expected from the fact that the system is now energy conserving.

For notational ease, the detailed balance property and 
$H$-theorem for DPD have been
presented for the single component case.  However, their extension to the
multi-component case can be achieved in a similar manner to the
isothermal $H$-theorem result presented here.

\section{Conclusions} We have shown that the desirable statistical
mechanical properties of
detailed balance and the existence of $H$-theorems may be extended to
general multi-component interacting DPD systems, whether maintained at
constant temperature or at constant energy. Of course, such properties
are rigorously valid in the continuous time limit, and are only 
approximately true for discrete-time implementations of these
algorithms. The approximations improve as the size of the time-step is
decreased. Detailed balance makes possible the theoretical analysis
of models based on the DPD equations of motion, while the $H$-theorem
provides a means to control numerical instabilities in computer simulations.  

\section*{Acknowledgements}

CAM acknowledges support from EPSRC (UK) and Unilever Research (UK).
PVC is grateful to the Department of Theoretical Physics and Wolfson
College, University of Oxford, for a Visiting Fellowship.


\section*{References}
\begin{thebibliography}{99}

\bibitem{McL}
McLeish T 1997 {\it Theoretical Challenges in the Dynamics of
    Complex Fluids} (NATO ASI Series, Kluwer, Dordrecht) vol~339

\bibitem{BAC}
Boghosian B M and Alexander F J 1997.  Proceedings of the
    1996 Conference on Discrete Models for Fluid Mechanics,
    {\em Int. J. Mod. Phys. C} vol~8

\bibitem{HK}
Hoogerbrugge P J and Koelman J M V A 1992 {\it Europhys. Lett.}
  vol~19 p~155

\bibitem{EW}
Espa\~nol P and Warren P 1995 {\it Europhys. Lett.}
  vol~30 p~191

\bibitem{colloids}
Koelman J M V A and Hoogerbrugge P J 1993 {\it Europhys. Lett.}
  vol~21 p~363;
Boek E S, Coveney P V and Lekkerkerker H N W 1996 
  {\it J.Phys.Cond.Matt.} Vol~8 p~9509

\bibitem{polymers}
Schlijper A G, Hoogerbrugge P J and Manke C W 1995 
  {\it J. Rheol.} vol~39 p~567
Madden W G, Kong Y, Manke C W and Schlijper A G 1994
  {\it Int. J. Thermophysics} vol~15 p~1093

\bibitem{CN}
Coveney P V and Novik K E 1996 {\it Phys. Rev. E}
  vol~54 p~5134

\bibitem{NC}  
Novik K E and Coveney P V 1997 {\it Int. J. Mod. Phys. C}
  vol~8 p~909

\bibitem{MBE1}
Marsh C A, Backx G and Ernst M H 1997 {\it Phys. Rev. E}
  vol~56 p~1676

\bibitem{MBE2}
Marsh C A, Backx G and Ernst M H 1997 {\it Europhys. Lett.}
  vol~38 p~411

\bibitem{MY}
Marsh C A and Yeomans J M 1997 {\it Europhys. Lett.}
  vol~37 p~511

\bibitem{energy}
Espa\~nol P 1997 {\it Europhys. Lett.}
  vol~40 p~631

\bibitem{DB}
Coveney P V and Espa\~nol P 1997 {\it J. Phys. A: Maths and General} 
  vol~30 p~779

\bibitem{RISKEN}
Risken H 1989 {\it The Fokker-Planck Equation} (Berlin: Springer) 

\end{thebibliography}
\end{document}